\documentclass[12pt]{article}  
\usepackage{graphicx}

\def\R{ {\rm R \kern -.31cm I \kern .15cm}}
\def\C{ {\rm C \kern -.15cm \vrule width.5pt \kern .12cm}}
\def\Z{ {\rm Z \kern -.27cm \angle \kern .02cm}}
\def\N{ {\rm N \kern -.26cm \vrule width.4pt \kern .10cm}}
\def\1{{\rm 1\mskip-4.5mu l} }
\def\lsim{\raise0.3ex\hbox{$<$\kern-0.75em\raise-1.1ex\hbox{$\sim$}}}
\def\gsim{\raise0.3ex\hbox{$>$\kern-0.75em\raise-1.1ex\hbox{$\sim$}}}
\def\noi{\noindent}

\def\beq{\begin{equation}}   \def\eeq{\end{equation}}
\def\bea{\begin{eqnarray}}  \def\eea{\end{eqnarray}}
\def\nn{\nonumber}
\def\noi{\noindent}
\def\beeq{\begin{eqnarray}} \def\eeeq{\end{eqnarray}}

\begin{document}

\begin{center} {\large \bf New Classes of Potentials for which} \vskip 2 truemm

 {\large \bf the Radial Schr\"odinger Equation} \vskip 2 truemm
 {\large \bf can be solved at Zero Energy}

 \par \vskip 5 truemm

{\bf Khosrow Chadan}\\ {\it Laboratoire de Physique
Th\'eorique}\footnote{Unit\'e Mixte de Recherche UMR 8627 - CNRS\par \hskip 0.4 truecm e-mail : Khosrow.Chadan@th.u-psud.fr}\\   
{\it Universit\'e de Paris XI, B\^atiment 210, 91405 Orsay Cedex,
France} \par \vskip 5 truemm

{\bf Reido Kobayashi} \\ {\it
Department of Mathematics}\\ {\it  Tokyo University of Science, Noda,
Chiba 278-8510, Japan} \par \vskip 2 truemm
\end{center}
\vskip 1 truecm
\centerline{\it Dedicated to Professor Shinsho Oryu for his sixty fifth anniversary}
\vskip 1 truecm 
\begin{abstract}
Given two spherically symmetric and short range potentials $V_0$ and
$V_1$ for which the radial Schr\"odinger equation can be solved
explicitely at zero energy, we show how to construct a new potential
$V$ for which the radial equation can again be solved explicitely at
zero energy. The new potential and its corresponding wave function are
given explicitely in terms of $V_0$ and $V_1$, and their corresponding
wave functions $\varphi_0$ and $\varphi_1$. $V_0$ must be such that it
sustains no bound states (either repulsive, or attractive but weak).
However, $V_1$ can sustain any (finite) number of bound states. The new
potential $V$ has the same number of bound states, by construction, but
the corresponding (negative) energies are, of course, different. Once this is
achieved, one can start then from $V_0$ and $V$, and construct a new
potential $\overline{V}$ for which the radial equation is again
solvable explicitely. And the process can be repeated indefinitely. We
exhibit first the construction, and the proof of its validity, for
regular short range potentials, i.e. those for which $rV_0(r)$ and
$rV_1(r)$ are $L^1$ at the origin. It is then seen that the
construction extends automatically to potentials which are singular at
$r= 0$. It can also be extended to $V_0$ long range (Coulomb, etc.). We give finally several explicit
examples.
\end{abstract}

\begin{flushleft}
LPT Orsay  05-69\\
October 2005
\end{flushleft}
 
\newpage
\pagestyle{plain}
\baselineskip 20pt
\noi {\bf I. Introduction} \\

Consider the reduced radial Schr\"odinger equation for a spherically
symmetric potential $V(r)$ \cite{1r}

\beq \label{1e}
\left \{ \begin{array}{l} \varphi ''_{\ell} (k, r) + k^2 \varphi_{\ell}
(k, r) = \left [ \displaystyle{{\ell (\ell + 1) \over r^2}} + V(r) \right
] \varphi_{\ell} (k, r) \ , \\ \\ r \in [0, \infty )\ , \ k \geq 0\ , \
\varphi_{\ell}(k, 0) = 0 \ , \\ \\ V(r) \ \hbox{real}\ , \
\hbox{locally} \ L^1\ \hbox{for}\ r \not = 0\ , \ \hbox{and}\ V(\infty
) = 0 \ .\end{array} \right .
  \eeq
\vskip 5 truemm 
  
\noi We exclude therefore confining potentials like the harmonic
oscillator, etc. \par

There are only a few potentials for which the radial Schr\"odinger
equation can be solved explicitely for all $k$ and all $\ell$. These
are essentially the square well, the Coulomb potential, sums of
$\delta$-function potentials \cite{1r}~; and for potentials which are
more singular than $r^{-2}$ at the origin, but repulsive, only $\lambda
/r^4$ \cite{2r}, for which the solution is given in terms of very
complicated Mathieu function.\par

If we restrict ourselves to the case of one single $\ell$, then we can
include the Bargmann potentials \cite{1r,3r}, for which, by
construction, the radial Schr\"odinger equation can be solved for one
specific value of $\ell$, and for all $k$. We remind the reader that
Bargmann potentials are those for which the $S$-matrix $S_{\ell}(k)$ is
a meromorphic function of $k$ in the $k$-plane. They can be constructed for each $\ell$.\par

In the particular case of $\ell = 0$, the radial equation can be solved
for all $k$ for the following potentials \cite{1r}~:

\beq
\label{2e}
V_1(r) = \lambda e^{-\mu r} \ ,
\eeq  

\beq
\label{3e}
V_2(r) = { \lambda e^{-\mu r} \over 1 - e^{-\mu r}} \quad \hbox{(Hulth\'en)} \ , 
\eeq

\vskip 5 truemm
\noi and, more generally, the Eckart potentials \cite{4r}~:

\beq
\label{4e}
V_3(r) = {\lambda_1 \ e^{-\mu r} \over 1 + C \ e^{-\mu r}} + {\lambda_2\ e^{-\mu r} \over \left ( 1 + C\ e^{-\mu r}\right )^2}\ ,
\eeq

\noi of which (\ref{3e}) is a particular case. The solutions are given in terms of hypergeometric functions. \\

In the case of $k = 0$, and for all $\ell$, one can add the potential \cite{5r}~:

\beq
\label{5e}
V_4(r) = {\lambda\ r^{\alpha - 2} \over \left ( C + r^{\alpha}\right )^2}\ , \quad \alpha > 0 \ , \ \ell \geq 0\ . 
\eeq
\vskip 5 truemm

\noi An interesting particular case is when $\alpha = 2$~:

\beq
\label{6e}
V_5(r) = {\lambda \over \left ( C + r^2\right )^2}\ , \quad \ell \geq 0\ .
\eeq
\vskip 5 truemm

Finally, for $k = 0$ and $\ell = 0$, one can solve the radial equation also for

\beq
\label{7e}
V_6(r) = {\lambda \over (C + r)^4}\ , \quad \ell = 0\ .
\eeq

\vskip 5 truemm
\noi We shall give later the explicit solutions for some of  these potentials, when they are simple.\par

The purpose of the present paper is to show that if one can solve
explicitely the radial Schr\"odinger equation at $k = 0$ for

\beq
\label{8e}
\left \{ \begin{array}{l} V_0(r) + \displaystyle{{\ell (\ell + 1) \over r^2}}\ , \\ \\ \displaystyle{\int_0^1} r |V_0 (r) \ dr < \infty \ , \ \displaystyle{\int_1^{\infty}} r^{2\ell + 2} |V_0(r) | dr < \infty \ ,\\ \\ \hbox{no bound states},\end{array} \right .
\eeq
\vskip 5 truemm
  
\noi and for $k = 0$ and $\ell = 0$ for the potential 

\beq
\label{9e}
\left \{ \begin{array}{l} V_1(r)\ , \ \ell = 0\ , \\ \\ \displaystyle{\int_0^1 r |V_1 (r) dr < \infty \ , \ \int_1^{\infty} r^2 |V_1(r) | dr < \infty} \ ,\\ \\ \hbox{any number of bound states},\end{array} \right .
\eeq
\vskip 5 truemm

\noi then one can solve explicitely the radial equation, again at $k =
0$ and $\ell = 0$, for the potential

\beq
\label{10e}
V(r) = V_0 (r) + F_0(r) \ V_1(G_0(r))\ ,
\eeq  

\vskip 5 truemm  
\noi in terms of the solutions for (\ref{8e}) and (\ref{9e}). Here, the
functions $F_0(r)$ and $G_0(r)$ are given very simply in terms of the
solutions for (\ref{8e}). \par

To begin with, we consider the case $\ell = 0$. If we call by $\varphi_0(r)$ and
$\chi_0(r)$ the two independent solutions for $V_0(r)$ defined by

\beq
\label{11e}
\left \{ \begin{array}{l} \varphi_0(0) = 0 \ , \ \varphi '_0(0) = 1\ ; \ \chi_0 (0) = 1 \ , \\ \\ W(\varphi_0, \chi_0) = \varphi'_0 \chi_0 - \varphi_0 \chi '_0 = 1 \ ,\end{array} \right .
\eeq  
 
 \vskip 5 truemm 
\noi $F_0(r)$ and $G_0(r)$ are given by

\beq
\label{12e}
F_0(r) = \left [ \chi_0 (r) \right ]^{-4} \ , \ G_0(r) = {\varphi_0 (r) \over \chi_0 (r)} \ .
\eeq
\vskip 5 truemm

\noi As we shall see later, since by assumption, $V_0(r)$ has no bound
states, all the above quantities are meaningful because both
$\varphi_0(r)$, and $\chi_0(r)$ defined by

\beq
\label{13e}
\chi_0 (r) = \varphi_0(r) \int_r^{\infty} {dt \over \varphi_0^2(t)} \ ,
\eeq
\vskip 5 truemm

\noi do not vanish anywhere, except for $\varphi_0$ at $r = 0$ (see details in the next section). Note that the converse of (\ref{13e}) is 

$$\varphi_0 (r) = \chi_0 (r) \int_0^r {dt \over \chi_0^2(t)} dt\ . \eqno(13')$$
\vskip 5 truemm

The solution of the radial equation at $k = 0$ and $\ell = 0$ for the
potential $V(r)$, (\ref{10e}), is then given by

\beq
\label{14e}
\left \{ \begin{array}{l} \varphi (r) = \chi_0 (r) \ \varphi_1 \left ( \displaystyle{{\varphi_0 (r) \over \chi_0 (r)}}\right ) \ , \\ \\ \varphi (0) = 0\ , \end{array} \right .
\eeq

\vskip 5 truemm

\noi where $\varphi_1$ is the regular solution of the radial equation
for $V_1$, (\ref{9e}), defined by $\varphi_1(0) = 0$. By assumption,
both $\varphi_0$ and $\varphi_1$ are known, and both vanish at $r = 0$
by definition. The above formula can be checked directly by
differentiation. We shall see in the next section how it was found.\\

\noi{\bf Remark 1.} As we shall see, $x = {\varphi_0(r) \over
\chi_0(r)}$ maps $r \in [0, \infty )$ into $x \in [0, \infty )$. The
mapping is one to one because of (\ref{39e}) below, and is, of course, twice differentiable (see (\ref{41e})).\\

Once (\ref{14e}) is known, it is easy to generalize it to the case one
has angular moment with $V_0(r)$~:

$$\left \{ \begin{array}{l} V_0(r) + \displaystyle{{\ell (\ell + 1) \over r^2}}\ , \quad \ell \geq 0\ , \\ \\ rV_0(r) \in L^1(0) \ \hbox{and}\ r^{2\ell+2} V_0(r) \in L^1(\infty )\ ,\\ \\ \hbox{no bound states},\end{array} \right . \eqno(8')$$  
  
\noi $V_1(r)$ remaining unchanged. $\varphi_0$ and $\chi_0$ are now the
solutions of the radial equation with the potentials ($8'$), so that (\ref{10e}) becomes

$$V(r) = \left [ V_0(r) + {\ell (\ell + 1) \over r^2}\right ] + F_0(r) V_1(G_0(r))\ . \eqno(10')$$

\noi All other formula given above remain unchanged. \par

In short, if one can solve explicitely the radial equations
  
\beq
\label{15e}
\left \{ \begin{array}{l} \varphi''_0(r) = \left [ V_0(r) + \displaystyle{{\ell ( \ell + 1) \over r^2}} \right ] \varphi_0 (r) \ , \quad \varphi_0(0) = 0 \ , \\ \\ \varphi ''_1(r) = V_1(r) \ \varphi_1(r) \ , \ \varphi_1 (0) = 0 \ , \end{array} \right .
\eeq 

\vskip 5 truemm  
\noi where $V_0$ and $V_1$ satisfy the assumptions shown in $(8')$
and (\ref{9e}), then the solution of

\beq
\label{16e}
\varphi '' (r) = \left [ V_0(r) + {\ell (\ell + 1) \over r^2} \right ] \varphi (r) + {1 \over \chi_0^4(r)} V_1 \left ( { \varphi_0 (r) \over \chi_0 (r)}\right ) \varphi (r) \ ,
\eeq
  
\noi with $\varphi (0) = 0$, is given by

\beq
\label{17e}
\varphi (r) = \chi_0 (r)\ \varphi_1 \left ( {\varphi_0 (r) \over \chi_0 (r)} \right ) \ .
\eeq
\vskip 5 truemm

\noi Remember that $\chi_0 (r)$ is always defined by (13).\par

It is easy to check our assertion by differentiating twice $\varphi$,
given by (\ref{17e}), and using (\ref{15e}). We shall see in the next
section how (\ref{15e}) and (\ref{17e}) were found. We shall also see
that one can replace $V_0(r)$ by strongly repulsive potentials which
are more singular than $r^{-2}$ at the origin. Examples will be
provided for

 \beq
 \label{18e}
 V_0(r) = {g \over r^n}\ , \quad g > 0\ , \quad n > 2 \ ,
 \eeq 
  
\noi for which the radial Schr\"odinger equation is soluble for all
$\ell$ at $ k = 0$ \cite{6r}. \\

\noi {\bf Remark 2, the bound states.}  As is well-known, the nodal
theorem \cite{7r} asserts that the number of bound states of $V(r)$,
(\ref{10e}) or (10$'$), is given by the number of the nodes of the
regular wave function $\varphi (r)$, (\ref{17e}). Since neither $\chi_0
(r)$ for $r \geq 0$, nor $\varphi_0 (r)$ for $r > 0$, do not vanish
(remember that, by assumption, $V_0$ has no bound states), and $x =
{\varphi_0 (r) \over \chi_0 (r)}$ maps $r \in [0, \infty )$ into $x \in
[0, \infty )$ and the mapping is one to one, it is obvious on
(\ref{17e}) that $\varphi$ and $\varphi_1$ have the same number of
nodes. Therefore, $V_1$, (\ref{9e}), and $V$, (\ref{10e}) or (10$'$),
have the same number of bound states. Of course, the energies of these
states are different for $V_1$ and $V$. In any case, one has also the
Bargmann bound for the number of bound states \cite{1r,3r,4r}~:

\beq
\label{19e}
n(V) = n(V_1) \leq \int_0^{\infty} r |V_1(r)|dr < \infty \ .
\eeq
\vskip 5 truemm

\noi {\bf Remark 3, iterating the process.} Once we have the explicit
solution (\ref{17e}) for the equation (\ref{16e}), we can start now
with the couple $[V_0(r), V(r)]$, instead of $[V_0(r) , V_1(r)]$, and
look for the solution of the radial equation at $k = 0$ for

\beq
\label{20e}
\overline{V}(r) = \left [ V_0(r) + {\ell (\ell + 1) \over r^2}\right ] + {1 \over \chi^4_0(r)} V \left [ {\varphi_0 (r) \over \chi_0 (r)}\right ] \ .
\eeq

\noi We will find now, of course, the solution

\beq
\label{21e}
\overline{\varphi}(r) = \chi_0 (r) \ \varphi \left ( {\varphi_0 (r) \over \chi_0 (r)} \right ) \ .
\eeq

\noi And this process can be continued as many times as we wish. \\

We end this introduction by giving one example with the potentials

\beq
\label{22e}
\left \{ \begin{array}{l} V_0(r) = \displaystyle{{\lambda a^2\over (1 + ar)^4}}\ , \quad a > 0\ , \ \lambda >0\ , \  \ell = 0\ , \\ \\ \varphi_0(r) = \left ( \displaystyle{{ 1 + ar \over a\sqrt{\lambda}}}\right ) \sinh \left ( \displaystyle{{\sqrt{\lambda}\ ar \over 1 + ar}}\right )\ , \\ \\ \chi_0 (r) =  (1 + ar) \left [ \cosh \left ( \displaystyle{{\sqrt{\lambda} \ ar \over 1 + ar}}\right ) - \displaystyle{{\cosh \sqrt{\lambda} \over \sinh \sqrt{\lambda}}}\sinh \left ( \displaystyle{{\sqrt{\lambda} \ ar  \over 1 + ar}}\right ) \right ] \end{array} \right .
\eeq

\noi and

\beq
\label{23e}
\left \{ \begin{array}{l} V_1(r) = \displaystyle{{g b^2 \over (b^2 + r^2)^2}}\ , \quad g > 0\ , \quad b > 0\ , \\ \\ \varphi_1(r) = \displaystyle{( b^2 + r^2 )^{1/2} \over \sqrt{g-1}}\sinh \left (  \sqrt{g-1}  \ {\rm Arctg}\  \displaystyle{{r \over b}}\right ) \end{array} \right .
\eeq

\vskip 5 truemm
\noi from which one can calculate $\varphi$ by formula (\ref{17e}).
Now, since the Schr\"odinger equation can be solved for the potential
(\ref{23e}) for all $\ell$, we can invert the roles of $V_0$ and $V_1$,
and start with

\beq
\label{24e}
V_0(r) + {\ell ( \ell + 1) \over r^2} = {g b^2 \over (b^2 + r^2)^2} + {\ell (\ell + 1) \over r^2} \ .
\eeq
 
\noi For general $\ell$,  the solutions $\varphi_0$ and $\chi_0$ are given in terms of hypergeometric functions \cite{5r}. We restrict ourselves to the case of $\ell = 0$, for which

\beq
\label{25e}
\left\{ \begin{array}{l} \varphi_0 (r) = \displaystyle{{\sqrt{b^2 + r^2}\over \sqrt{g-1}}} \sinh \left ( \sqrt{g-1} \ {\rm Arctg}\ \displaystyle{{r \over b}}\right ) \ ,\\ \\ \chi_0 (r) = \sqrt{b^2 + r^2} \cosh \left (\sqrt{g-1}\  {\rm Arctg}\ \displaystyle{{r \over b}} \right )  \ ,\end{array} \right . 
\eeq 
 \vskip 5 truemm
 
\noi and take (\ref{22e}) as the second potential, with

\beq
\label{26e}
\varphi_1(r) = {1 + ar \over a \sqrt{\lambda}} \sinh \left ( {\sqrt{\lambda} \ ar \over 1 + ar}\right ) \ .
\eeq 
 
 \vskip 8 truemm

\noi {\bf II. Derivation of the solution (\ref{14e})}\\

We begin by studying the properties of the solutions $\varphi_0 (r)$,
and $\chi_0 (r)$ defined by (\ref{13e}), of the radial equation at $k =
0$ and $\ell = 0$~:

\beq
\label{27e}
\left \{ \begin{array}{l} \varphi''_0 (r) = V_0 (r)\ \varphi_0 (r) \ ,\\ \\ \varphi_0 (0) = 0\ , \ \chi_0 (0) = 1 \ ,\end{array} \right .
\eeq

\vskip 5 truemm
\noi where $V_0(r)$ satisfies the assumptions shown in (\ref{8e}). Since
the radial equation is homogeneous, we can normalize its solution
$\varphi$ as we wish. For (\ref{27e}), the usual convention is to put

\beq
\label{28e}
\varphi'_0 (0) = 1 \ .
\eeq

 \vskip 5 truemm
 \noi Then we can combine (\ref{27e}) and (\ref{28e}) into a single
Volterra integral equation \cite{1r,3r,4r}

\beq
\label{29e}
\varphi_0 (r) = r + \int_0^r (r - r') V_0(r') \varphi_0 (r') dr' \ .
\eeq

 \vskip 5 truemm
\noi It can then be shown that, iterating the above equation, and using
the assumptions on $V_0 (r)$, namely that $rV$ is $L^1$ at $r = 0$,
$r^2V(r)$ is $L^1$ at $r = \infty$, one gets an absolutely and
uniformly convergent series defining the solution $\varphi_0 (r)$,
together with the bound \cite{1r,3r,4r}

\beq
\label{30e}
|\varphi_0(r)| \leq r\ e^{\int_0^r r'|V_0(r')|dr'} < Cr\ ,
\eeq

\vskip 5 truemm
\noi where $C$ is an absolute constant less than $\exp \int_0^{\infty}
r |V_0(r)|dr$. Using this bound in (\ref{29e}), we find that indeed,
for $r \to 0$, we have

\beq
\label{31e}
\varphi_0 (0) = 0\quad , \quad \varphi '_0 (0) = 1 \ ,
\eeq

\noi and for $r \to \infty$,

\beq
\label{32e}
\varphi_0(r) = r \left [ 1 + \int_0^{\infty} V_0 (r') \varphi_0 (r') dr' \right ] - \int_0^{\infty} r' V_0 (r') \varphi_0 (r') dr' + o(1) \ .
\eeq

\noi where all integrals are absolutely convergent.\par

There are now two cases~: \par

{\bf i)} the potential $V_0(r)$ is positive. Then it is obvious on the
iterated series of (\ref{29e}) that all the terms are positive, and so
is $\varphi_0 (r)$ for all $r$. It follows then from (\ref{27e}) that
$\varphi_0 (r)$ is a positive convex function of $r$. It increases
indefinitely, and we have, on the basis of (\ref{32e})~:

\beq
\label{33e}
\left \{ \begin{array}{l} \varphi_0(r) > 0\ , \ \hbox{and convex},\\ \\ \varphi_0 (r) = Ar + B + o(1)\ , \ \hbox{as}\ r \to \infty\ , \ 1 < A < \infty\ , \ B < 0 \ . \end{array} \right .
\eeq

\vskip 5 truemm
\noi A schematic picture of $\varphi_0$ is shown on Fig. 1. \par

{\bf ii)} $V_0 (r) < 0$, but not strong enough to have bound states.
Then we find from the nodal theorem relating the bound states of $V_0
(r)$ to the zeros (nodes) of $\varphi_0 (r)$ for $r > 0$ \cite{7r},
that $\varphi_0(r)$ is again positive, and since $V_0$ is now negative,
$\varphi_0$ is positive and concave. A schematic picture of $\varphi_0$
is shown on Fig. 1. From (\ref{32e}), (\ref{33e}) is now replaced by

\beq
\label{34e}
\left \{ \begin{array}{l} \varphi_0 (r) > 0\ , \ \hbox{and concave},\\ \\ \varphi_0 (r) = Ar + B + o(1)\ , \ \hbox{as} \ r \to \infty\ , \ 0 < A < 1\ , \ B > 0\ . \end{array} \right .
\eeq

\vskip 5 truemm

\noi {\bf Remark 4.} Here, if $A < 0$, then, since $\varphi '_0(0) =
1$, and $\varphi '_0 (\infty ) < 0$, $\varphi_0$ must have a zero in
between, and therefore one has a bound state, in contradiction with our
assumption of no bound states. If $A = 0$, this means that one is at
the threshold of having a bound state. More precisely, that one has a
resonance at zero energy \cite{1r,3r,4r}, a possibility we have
excluded also.\\

We come now to the second, independent solution $\chi_0(r)$, defined by
(\ref{13e}). First of all, since $\varphi_0 (r)$ is always positive for
$r > 0$, and from (\ref{33e}) and (\ref{34e}), the integral is
absolutely convergent at its upper limit, and so $\chi_0$ is twice
differentiable, and satisfies the same equation as $\varphi_0$. It is
trivial to show that the Wronskian of the two, (\ref{11e}), is 1. Now,
when $r \to 0$, the integral in (\ref{13e}) diverges, but since
$\varphi_0 (r) = r + o(1)$ as $r \to 0$, and there is $\varphi_0(r)$ in
front of the integral, it is trivial to show that we have $\chi_0 (0) =
1$, as shown in (\ref{11e}). Also, on the basis of the Wronskian
(\ref{11e}), and (\ref{31e}), we find

\beq
\label{35e}
\lim_{r\to 0} \ r \ \chi'_0(r) = 0 \ .
\eeq

\vskip 5 truemm
\noi This general property is a consequence of $rV_0(r) \in L^1$ at the
origin. The derivative of $\chi_0 (r)$ at $r = 0$ maybe finite or
infinite, depending on the behavior of $V_0(r)$ near $r = 0$. If $V_0
(r)$ itself is $L^1$ at $r = 0$, one can write also the integral
equation \cite{1r,3r,4r}~:

\beq
\label{36e}
\chi_0(r) = 1 + \int_0^r (r - r') V_0(r') \chi_0 (r') dr' \ ,
\eeq

\vskip 5 truemm
\noi and iterate it, as we did with (\ref{29e}) for $\varphi_0$, to
find the solution, which turns out now to be bounded everywhere. One
then immediately sees on (\ref{36e}) that $\chi'_0(0)$ is finite. We
have, therefore, according to (\ref{13e}), and (\ref{33e}) or
(\ref{34e}) (see Fig. 2) : 

\beq
\label{37e}
\left\{ \begin{array}{l} \chi_0(0) = 1\ , \quad \chi_0 (r) > 0 \ \hbox{for all $r$}\ ,\\ \\ \chi_0 (r) \ \hbox{is a convex and decreasing function when $V_0 > 0$}\ ,\\ \\ \chi_0(r)\ \hbox{is a concave and increasing function when $V_0 < 0$,}\\
\hbox{with no bound states},\\ \\ \chi_0 (\infty) = \displaystyle{{1 \over A}} \not= 0 ,  \infty\ ; \ V_0 > 0 \Rightarrow A > 1\ ,\ V_0 < 0 \Rightarrow A < 1\ . \end{array} \right .
\eeq

\vskip 5 truemm
Consider now the mapping~:

\beq
\label{38e}
r \to x(r) = {\varphi_0 (r) \over \chi_0 (r)} \ .
\eeq

\noi According to (\ref{37e}), this a perfectly regular and
differentiable mapping, and is one to one since, according to (\ref{11e}), we have

\beq
\label{39e}
{dx \over dr} = {\varphi '_0 \chi_0 - \varphi_0 \chi '_0 \over \chi_0^2(r)} = {1 \over \chi_0^2(r)} > 0 \ .
\eeq 
 
\noi It follows then, since $\varphi_0 (r \to \infty ) \to \infty$, and
$\chi_0 (r \to \infty ) \to {1 \over A} \not= 0, \infty$, that the
mapping is one to one~:

\beq
\label{40e}
r \in [0, \infty ) \Leftrightarrow x \in [0, \infty )\ , \ x(0) = 0 \ , \ x(\infty ) = \infty \ .
\eeq 
 
\noi In fact, this mapping is twice continuously differentiable since

\beq
\label{41e}
{d^2x \over dr^2} = {- 2 \chi '_0 (r) \over \chi_0^3(r)}\ ,
\eeq

\vskip 5 truemm
\noi and $\chi '_0 (r)$ is a continuous function of $r$ for $r \geq 0$.
This last property follows from $\chi ''_0 (r) = V_0(r) \chi_0 (r)$,
where, by assumption, $V_0(r) \in L^1$ for $r > 0$. Since $\chi_0 (r)$
is a continuous function, $\chi ''_0 (r)$ is also $L^1$ for all $r >
0$, and so $\chi '_0 (r)$ is continuous for $r > 0$. $\chi '_0$ cannot
have jumps \cite{9r}. See Fig. 3.\par

Once we have established that the mapping (\ref{40e}) is regular and
twice continuously differentiable, we can consider the equation

\beq
\label{42e}
\left \{ \begin{array}{l} \varphi ''(r) = \left [ V_0 (r) + V_1 (r) \right ] \varphi (r)\ , \\ \\ \varphi (0) = 0 \ , \ \varphi ' (0) = 1\ ,\end{array} \right .
\eeq
 
 \vskip 5 truemm
\noi where $V_0$ and $V_1$ satisfy the conditions shown in (\ref{8e})
and (\ref{9e}). We make now the change of variable and function

\beq
\label{43e}
\left . r \to x = {\varphi_0 (r) \over \chi_0 (r)}\ , \quad \psi (x) = {\varphi (r) \over \chi_0 (r)} \right |_{r = r(x)}\ ,
\eeq 
 
 \vskip 5 truemm
\noi where $r(x)$ is the inverse mapping, i.e. the inverse function of
$x = x(r)$. Obviously, $r(x)$ is also twice continuously
differentiable. Differentiating now twice $\psi$ with respect to $x$,
and using (\ref{13e}), we easily find

\beq
\label{44e}
\left . \ddot{\psi} (x) = \left [ \chi_0^4 (r) V_1(r) \right ] \right |_{r= r(x)} \psi (x) \ .
\eeq 
 
 \vskip 5 truemm
\noi There is no longer $V_0$ present. From the definition of $\psi
(x)$, of $\varphi (r)$ given in (\ref{42e}), and (\ref{35e}), it is
obvious that, because $\varphi (r) = r + o(1)$ as $r \to 0$, we have 

\beq
\label{45e}
\psi (0) = 0\ , \ \dot{\psi} (0) = \lim_{x \to 0} \dot{\psi} (x) = \lim_{r\to 0} \left [ \varphi ' (r) \chi_0 (r) - \varphi (r) \chi '_0 (r) \right ] = 1 \ .
\eeq

\vskip 5 truemm
Suppose now that, from the beginning, $V_1(r)$ in (\ref{42e}) was of
the form,

\beq
\label{46e}
{1 \over \chi_0^4(r)} V_1\left ( x = {\varphi_0 (r) \over \chi_0 (r)}\right ) \ ,
\eeq

\vskip 5 truemm
\noi where $x V_1(x) \in L^1$ at $x = 0$, and $x^2V_1(x) \in L^1$ at $x
= \infty$. Then (\ref{44e}) would become

\beq
\label{47e}
\left \{ \begin{array}{l}  \ddot{\psi} (x) = V_1(x) \ \psi (x) \ ,\\\\ \psi (0) = 0\ , \ \dot{\psi} (0) = 0 \ , \end{array} \right .
\eeq

\vskip 5 truemm
\noi which we assume to be explicitely solvable. Then, from the
definition (\ref{43e}), we would have for the solution of (\ref{42e}),
with a $V_1$ of the form (\ref{46e}),

\beq
\label{48e}
\varphi (r) = \chi_0 (r) \ \psi \left ( {\varphi_0 (r) \over \chi_0 (r)}\right )\ .
\eeq

\vskip 5 truemm
\noi Combining all these, with a slightly different notation, we have
therefore the following \\

\noi {\bf Theorem 1.} The solution of

\beq
\label{49e}
\left \{ \begin{array}{l} \varphi '' (r) = \left [ V_0 (r) + \displaystyle{{1 \over \chi_0^4 (r)}} V_1 \left ( \displaystyle{{\varphi_0 (r) \over \chi_0 (r)}}\right ) \right ] \varphi (r)\\ \\ \varphi (0) = 0 \ ,\ \varphi ' (0) = 1\ , \end{array} \right .
\eeq

\vskip 5 truemm
\noi where $\varphi_0$ and $\chi_0$ are the two solutions of 

\beq
\label{50e}
\left \{ \begin{array}{l} \varphi ''_0 (r) = V_0 (r) \ \varphi_0 (r) \ , \\ \\ \varphi_0 (0) = 0 \ ,\ \varphi '_0 (0) = 1\ , \ \hbox{no bound states,}\\ \\ \chi_0(r) \ \hbox{defined by (\ref{13e})}, \chi_0 (0) = 1\ , \end{array} \right .
\eeq

\noi is given by

\beq
\label{51e}
\varphi (r) = \chi_0 (r)\ \varphi_1 \left ( {\varphi_0 (r) \over \chi_0 (r)}\right ) \ ,
\eeq

\noi where $\varphi_1 (x)$ is the (regular) solution of

\beq
\label{52e}
\left \{ \begin{array}{l} \ddot{\varphi}_1 (x) = V_1 (x) \ \varphi_1 (x) \ , \\ \\ \varphi_1 (0) = 0 \ ,\ \dot{\varphi}_1 (0) = 1\ . \end{array} \right .
\eeq

\vskip 5 truemm
\noi Therefore, if the Schr\"odinger equation at $k = 0$ and $\ell = 0$
can be explicitely solved for $V_0$ and $V_1$, then the solution of
(\ref{49e}) is of the form (\ref{51e}). As we said in the introduction,
one can check directly, that (\ref{51e}) is indeed the solution of
(\ref{49e}). For bound states in (\ref{52e}) and (\ref{49e}), see below, after Remark 6.\\

{\bf Higher waves, $\ell {\bf >}$ 0.} So far, we have been assuming
$\ell = 0$. It is easy to extend the results to the case $\ell > 0$ in
(\ref{50e}), i.e. to begin with

\beq
\label{53e}
\varphi ''_0 (r) = \left [ V_0 (r) + {\ell (\ell + 1) \over r^2} \right ] \varphi_0 (r) \ ,
\eeq

\vskip 5 truemm
\noi and then add the potential $V_1$ in (\ref{49e}). $V_0$ is, as before,
supposed to be such that $r V_0 \in L^1$ at $r = 0$, and $r^2V_0 \in
L^1$ at $r = \infty$. We shall see later that we need more rapid
decrease at infinity. The regular solution $\varphi_0$ is usually
normalized as follows

\beq
\label{54e}
\varphi_0 (r) = {r^{2\ell + 1} \over (2\ell + 1) !!} + o\left ( r^{\ell + 1} \right ) \ , \quad r \to 0 \ .
\eeq

\vskip 5 truemm
\noi One can then combine (\ref{53e}) and (\ref{54e}) into the single
Volterra integral equation

\beq
\label{55e}
\varphi_0 (r) = {r^{\ell + 1} \over (2\ell + 1)!!} + \int_0^r {r^{2 \ell + 1} - r'^{2\ell +1} \over (2\ell + 1)r^{\ell} r'^{\ell}} V_0 (r') \varphi_0 (r') dr'\ .
\eeq

\vskip 5 truemm
\noi Solving this equation by iteration, one finds again, as for the
case $\ell = 0$, an absolutely and uniformly convergent series defining
the solution $\varphi_0$, together with a bound similar to (\ref{30e})
for all finite $r \geq 0$~:

\beq
\label{56e}
|\varphi_0(r)| \leq C\ r^{\ell + 1} \exp \left ( \int_0^r r' |V_0(r)|dr'\right ) \leq C'\ r^{\ell + 1}\ ,
\eeq 

\vskip 5 truemm
\noi where $C$ and $C'$ are absolute finite constants \cite{1r,3r,4r}.
Using (\ref{56e}) in (\ref{55e}), one sees immediately that, for $r \to
0$,

\beq
\label{57e}
\varphi_0 (r) = {r^{\ell + 1} \over (2 \ell + 1)!!} + o\left ( r^{2\ell + 1}\right )\ , \ \varphi '_0 (r)= {r^{2\ell} \over (2 \ell - 1)!!} + o \left ( r^{2\ell }\right ) \ .  
\eeq

\vskip 5 truemm
\noi For all the above results, we need only $r V_0 \in L^1(0)$. Also,
by assumption, there are no bound states for (\ref{53e}). It follows
again that, by the nodal theorem \cite{7r}, $\varphi_0 (r)$ cannot
vanish for $r > 0$. Because of (\ref{57e}), we find therefore that

\beq
\label{58e}
\varphi_0 (r) > 0\ \hbox{for all $r > 0$} \ .
\eeq

\noi From this, and (\ref{53e}), it follows immediately that if $V_0(r)
> 0$, then $\varphi_0 (r)$ is convex. For $V_0 (r) < 0$, the situation
is more subtle than for the case of $\ell = 0$, nd $\varphi_0$ may
become concave in some interval $(R_1, R_2)$.\par

We wish now to look at the behaviour of $\varphi_0 (r)$ as $r \to \infty$. We assume here

\beq
\label{59e}
r^{2\ell + 2} \ V_0 (r) \in L^1 (\infty ) \ .
\eeq

\noi Then we can let $r \to \infty$ in (\ref{55e}), to find
\bea
\label{60e}
&&\varphi_0 \ \mathrel{\mathop {\rm =}_{r\to \infty}}\  \left [ {1 \over (2 \ell + 1)!!} + \int_0^{\infty} {1 \over 2 \ell + 1} r'^{- \ell} V_0 (r') \varphi_0(r') dr'\right ] r^{\ell + 1}  \nn \\
&&\qquad - {1 \over 2\ell + 1} \left [ \int_0^{\infty} r'^{\ell + 1} V_0 (r') \varphi_0 (r') dr' \right ] r^{-\ell} + \cdots \nn \\
&&\qquad = A_{\ell} \ r^{\ell + 1} + B_{\ell} \ r^{-\ell} + \cdots  
\eea

\noi Since $\varphi_0 (r)$ never vanishes (absence of bound states),
$A_{\ell}$ must be always positive. If $V_0 > 0$, then $A_{\ell} > 1$,
and if $V_0 < 0$, $0 < A_{\ell} < 1$. For $B_{\ell}$, it is just the
opposite. These are quite the same as for $\ell = 0$.\par

We can now define the second (independent) solution $\chi_0(r)$ by
(\ref{13e}) again, and we find now, using (\ref{57e}) and (\ref{60e}),
that

\beq
\label{61e}
\left \{ \begin{array}{l} \chi_0 (r) = \displaystyle{{(2 \ell - 1) !! \over r^{\ell}}} + \cdots \ , \quad r \to 0 \\ \\ \chi_0 (r) = \displaystyle{{1 \over (2 \ell + 1) A_{\ell}}} r^{-\ell}\ , \quad r \to \infty \ . 
\end{array} \right .
\eeq

\vskip 5 truemm
\noi If we introduce the same variable $x = x(r)$ as before

\beq
\label{62e}
x = x(r) = {\varphi_0 (r) \over \chi_0 (r)} \ ,
\eeq

\noi we find

\beq
\label{63e}
\left \{ \begin{array}{l} x(r) = A \ r^{2 \ell + 1} + \cdots\ , \quad r \to 0 \\ \\ x(r) = B\ r^{2\ell + 1} \ , \quad r \to \infty \ .
\end{array} \right . \eeq

\noi The rest of the analysis goes as before, and we find\\

\noi {\bf Theorem 2.} Theorem 1 is valid if we add $\ell (\ell +
1)/r^2$ to $V_0$ in (\ref{49e}) and (\ref{50e}), i.e.

\beq
\label{64e}
\left \{ \begin{array}{l} \varphi '' (r) = \left [ \left ( V_0 (r) + \displaystyle{{\ell ( \ell + 1) \over r^2}}\right )  + \displaystyle{{1 \over \chi_0^4(r)}} V_1 \left ( \displaystyle{{\varphi_0 (r) \over \chi_0 (r)}}\right ) \right ] \varphi (r)\ , \\ \\ \varphi (r \to 0) = \displaystyle{{r^{\ell + 1} \over ( 2 \ell + 1) !!}} + \cdots \end{array} \right .
\eeq

\noi and

\beq
\label{65e}
\left \{ \begin{array}{l} \varphi ''_0 (r) = \left [ V_0 (r) + \displaystyle{{\ell ( \ell + 1) \over r^2}} \right ] \varphi_0 (r)\ , \ r^{2\ell + 2} \ V_0(r) \in L^1(\infty )\ ,\\ \\ \varphi_0 (r \to 0) = \displaystyle{{r^{\ell + 1} \over ( 2 \ell + 1) !!}} + \cdots \ , \ \hbox{no bound states,}\end{array} \right .
\eeq

\vskip 5 truemm
\noi $\chi_0(r)$ defined by (\ref{13e}), $\chi_0 (r \to 0) = {(2\ell - 1)!! \over r^{\ell}}$, while (\ref{52e}) remains unchanged. The solution is again
provided by (\ref{51e}). This can also be checked directly.\\

\noi {\bf Remark 5.} Since the behaviour of $\varphi_1(x)$ is $x +
\cdots$ as $x \to 0$, and $\varphi_0 (r) /\chi_0 (r) = r^{2\ell + 1}$ as $r \to 0$,
it is obvious on (\ref{51e}) that we have, as we should,

\beq
\label{66e}
\varphi (r) = \alpha_{\ell} \ r^{\ell + 1} + \cdots \ , \quad r \to 0 \ .
\eeq

\noi Likewise, it is easy to find

\beq
\label{67e}
\varphi (r) = \beta_{\ell} \ r^{\ell + 1} + \gamma_{\ell} \ r^{-\ell} \ , \quad r \to \infty\ .
\eeq

\vskip 5 truemm

\noi {\bf Bound States.} So far, we have assumed that
there are no bound states in (\ref{52e}). If there are
$n$ bound states with $V_1(x)$, i.e. in (\ref{52e}), which is the same
for Theorem 1 and Theorem 2, this means that $\varphi_1$ has $n$ nodes
for $x > 0$. And we have also $n$ nodes for the full solution
$\varphi$, given always by (\ref{51e}). The potentials $V_1$ and $V$
have the same number of bound states. But, of course, the binding
energies are different. \\

\noi {\bf Remark 6.} As we said in Remark 3, once we have $V(r)$ and
$\varphi (r)$, we can now start again with $V_0$ and $V$ instead of
$V_0$ and $V_1$, and proceed as before. This process can be repeated as
many times as we wish, and we get more and more potentials for which
the radial Schr\"odinger equation at zero energy can be solved. Also,
the number of bound states, if any, remains the same. Unfortunately,
the potentials and the wave functions become quickly very complicated.
However, one may ask what will happen at the limit of infinite
repetitions. This question is certainly not easy to answer. \\

\noi {\bf Singular Potentials.} As we said in the introduction, the
radial equation can be solved at $k = 0$ and for all $\ell$ for singular
potentials which are just inverse powers potentials shown in
(\ref{18e}). The solutions are given in terms of modified Bessel and
Hankel functions. We shall see explicit examples in the next section.
One more class is given by \cite{6r,10r}

\beq
\label{68e}
V_0 (r) = {g_1 \over r^2\left ( {\rm Log}\ {1 \over r}\right )^p} + {g_2 \over r^2 \left ( {\rm Log}\ {1 \over r}\right )^2} + {\ell (\ell + 1) \over r^2}\ ,
\eeq 

\vskip 5 truemm
\noi with $p < 2$, $g_1 > 0$. These potentials must be cut at $r = R_0
< 1$ in order to avoid none integrable singularities at $r = 1$. The
solution is given in terms of Whittaker functions \cite{6r,10r}. Once
we know the solution $\varphi_0 (r)$, we can proceed as before, and add
a regular potential $V_1$ with any (finite) number of bound states.\par

We should mention here that, contrary to the case of regular potentials
at the origin, i.e. those for which $r V_0(r)\in L^1$ at $r= 0$, here,
because of strong singularities at $r = 0$, we find \cite{1r,6r}
$\varphi_0 (0) = \varphi '_0 (0) = \cdots \varphi_0^{(n)}(0) = \cdots =
0$. All the derivatives of $\varphi_0$ vanish at $r = 0$. The
normalization is therefore arbitrary, and cannot be made at $r = 0$.
Once this is chosen (usually by the behaviour of $\varphi$ at $r =
\infty$), then $\chi_0 (r)$ is given again by (\ref{13e}), and we have
the Wronskian $W(\varphi_0, \chi_0) = 1$. Usually, in such a case, it is customary to normalize $\varphi_0$ at infinity, according to

$$\varphi_0 (r) = r + C + o(1)\ , \ r \to \infty\ , \eqno({\rm 69a})$$

\noi which entails also

$$\chi_0 (r) = r + C' + o(1) \ , \ r \to \infty \ . \eqno({\rm 69b})$$

\noi As we shall see on explicit
examples in the next section, our procedure for generating new
potentials can go on without modifications.\\

\noi {\bf III. Examples}\\

\noi {\bf 1. Regular Potentials.} We have already given, in the
introduction, as examples for the applications of Theorems 1 and 2, the
solutions of the Schr\"odinger equation for $\ell = 0$, $V_0 = (7)$,
$V_1 = (6)$, or $V_0 = (6)$ together with $\ell (\ell + 1)/r^2$, and $V_1
= (7)$. They are given by formulae (\ref{22e})-(\ref{26e}). More
examples are obtained by combining any two potentials among those given
by (\ref{2e}) to (\ref{7e}). We need only the solutions $\varphi$ and
$\chi$, the latter defined by (\ref{13e}), for these potentials. \\

{\it a. Exponential potential}

$$
\left \{ \begin{array}{l} V(r) = \lambda e^{-\mu r}\ , \ \lambda > 0\ ,
\ \mu > 0\ , \ \ell = 0 \\ \\ \varphi (r) = \alpha \ I_0 \left (
\displaystyle{{2 \sqrt{\lambda} \over \mu}} e^{- \mu r/2}\right ) +
\beta \ K_0 \left ( \displaystyle{{2 \sqrt{\lambda} \over \mu}} e^{- \mu
r/2}\right )\ , \\ \\ \chi (r) = \displaystyle{{I_0 \left ( {2
\sqrt{\lambda} \over \mu} e^{-\mu r/2} \right ) \over I_0 \left ( {2
\sqrt{\lambda} \over \mu}\right)}}\ , \quad \chi (0) = 1\ , \quad \chi(\infty ) =
\displaystyle{{1 \over I_0 \left ( {2 \sqrt{\lambda} \over \mu}\right)
}} \ .\end{array} \right .
\eqno(70)$$

\vskip 5 truemm
\noi $I_0$ and $K_0$ are modified Bessel and Hankel functions of order
zero, and the constants $\alpha$ and $\beta$ are determined to have
$\varphi (0) = 0$ and $\varphi ' (0) = 1$. It is then easy to show
that, according to our general analysis of section II, we have

$$
\varphi (r \to \infty ) = I_0 \left ( {2 \sqrt{\lambda} \over \mu}\right ) r + \cdots 
\eqno(71)$$

\vskip 5 truemm
\noi The presence of $r$ in (71) is due to the presence of $\log
z$ in $K_0(z)$ where $z \to 0$ \cite{8r}.\\

{\it b. The potential (\ref{5e}) for $\alpha > 0$, $\ell \geq 0$.}
The solutions $\varphi$ and $\chi$ are given by combinations of
hypergeometric functions of appropriate  arguments. We refer the reader
to \cite{5r} for details.\\

{\it c. Hulth\'en Potential ($3$), $\ell = 0$.} Here also, the
solutions $\varphi$ and $\chi$ are given in terms of appropriate
hypergeometric functions $F$. See \cite{1r,4r} for details. \\

\noi {\bf 2. Singular Potentials \cite{6r}.} Here, we consider only three cases.\\

{\it d. Inverse Power Potentials, $\ell \geq 0$~:}

$$
\left \{ \begin{array}{l} V(r) = \displaystyle{{g \over r^n}}\ , \quad g
>0\ , \quad n > 2\ell + 3\ , \\ \\ \varphi (r) = r^{1/2} K_{{2 \ell + 1
\over n-2}} \left ( \displaystyle{{2\sqrt{g} \over
(n-2)r^{{n-2\over2}}}}\right ) \ , \\ \chi (r) = r^{1/2} \left [ \alpha
K_{{2\ell + 1 \over n-2}} \left ( \displaystyle{{2\sqrt{g} \over (n-2)
r^{{n-2 \over 2}}}}\right ) + \beta I_{{2\ell + 1 \over n- 2}} \left (
\displaystyle{{2\sqrt{g} \over (n-2) r^{{n-2 \over 2}}}}\right ) \right .\
,\end{array} \right .
\eqno(72)$$

\vskip 5 truemm
\noi where $I_{\nu}$ and $K_{\nu}$ are modified Bessel and Hankel
functions. We must choose $n > 2\ell + 3$ in order to comply with
(\ref{60e}) \cite{6r}. \par

The parameters $\alpha$ and $\beta$ are determined for having $\chi
(r)$ to comply with the asymptotic behaviours deduced from (\ref{13e}),
for $r \to 0$ and $r \to \infty$. One has to remember the Wronkian

$$W\left [ r^{1/2} \ K_{\nu}\left ( \beta r^{\sigma}\right ) , r^{1/2} I_{\nu}\left ( \beta r^{\sigma}\right ) \right ] = \sigma\ . \eqno(73)$$

The case $\ell = 0$, $n = 4$ is particulary simple. One finds

$$
\left \{ \begin{array}{l} V(r) = \displaystyle{{g \over r^4}}\ , \quad g
>0\ , \quad \ell = 0\ , \\ \\ \varphi (r) = r e^{-\sqrt{g}/r} \
\displaystyle{\mathrel{\mathop {\rm =}_{r\to \infty}}}\ r - \sqrt{g} +
\cdots \\ \\ \chi (r) = \displaystyle{{r \over \sqrt{g}}} \sinh \left
( \displaystyle{{\sqrt{g}\over r }}\right ) \
\displaystyle{\mathrel{\mathop {\rm =}_{r\to \infty}}}\ 1 +
\cdots\end{array}\right .
\eqno(74)$$

\vskip 5 truemm
{\it e. Logarithmic Potentials} \cite{6r,10r}. We consider here the
simplest case of (\ref{68e}) with $p = 1$, and any angular momentum
$\ell \geq 0$~:

$$
\left \{ \begin{array}{l} V(r) = \left [ \displaystyle{{\alpha \over
r^2\ {\rm Log} \left ({1 \over r}\right )}} + \displaystyle{{g \over
r^2\ {\rm Log}^2\ \left ( {1 \over r}\right)}}\right ] \theta (R- r)\ ,
\ \alpha > 0\ , \ R < 1\ ,\\ \\ \varphi (r) = r^{1/2} \left [ \displaystyle{{\Gamma (- 2 \nu ) \over \Gamma \left ( {1 \over 2} - \nu - \mu\right )}} \ M_{\mu , \nu}(x)\right .\\ \qquad \left . + \displaystyle{{\Gamma (2 \nu ) \over \Gamma \left ( {1 \over 2} + \nu - \mu \right )}}\ M_{\mu, - \nu}(x)\right ] \ ,\\ \\ \chi (r) = r^{1/2} \left [ \alpha \ M_{\mu , \nu}(x) + \beta M_{\mu , - \nu}(x) \right ] \ ,   
\end{array}\right .
\eqno(75)$$

\vskip 5 truemm

\noi where $M_{\mu , \nu}$ are Whittaker functions \cite{8r},

$$x = (2 \ell + 1) {\rm Log}\ {1 \over r}\ ,\quad k = {- \alpha \over 2 \ell + 1}\ , \quad \nu = i \sqrt{g - {1 \over 4}}\ , \eqno(76)$$

\noi and $\alpha$ and $\beta$ are determined according to (\ref{13e}) for $r \to 0$ and $r \to \infty$. Note here that, for $r \geq R$, we have the free equation (no potential), and, therefore, we must first adjust the free solution to the interior solution at $r = R$, as usual.\par

 Note here that the singular part of the potential is just the
first potential, and that is why we must choose $\alpha > 0$. The
second potential is regular since it satisfies $r V(r) \in L^1(0)$. We
can therefore choose $g \ {{\textstyle >}\atop{\textstyle <}}\ 0$.
There are several more examples of singular potentials for which the
radial equation can be solved explicitely. We refer the reader for
details to \cite{6r}.\\

{\it f. Coulomb Potential.} The Coulomb potential is regular at the origin, and so we have the usual solutions $\varphi$ and $\chi$, as defined previously. We choose, of course, the repulsive case. The solutions can be read off from (72) adapted to $n < 2$, or else, be obtained in the standard way \cite{1r,4r}. One has, for $\ell = 0$~: 

$$\left \{ \begin{array}{l} V = \displaystyle{{\alpha \over r}} \ , \quad \alpha > 0 \ , \quad \ell = 0 \ , \\ \\ \varphi (r) = \displaystyle{\sqrt{{r \over \alpha}}} \ I_1(2 \sqrt{\alpha r})\ , \\ \\ \chi (r) = - \pi \sqrt{\alpha r} \ K_1 \left ( 2 \sqrt{\alpha r}\right ) \ ,\end{array}\right . \eqno(72')$$

\noi where $I_1$ and $K_1$ are the modified Bessel and Hankel functions, and similar formulae for $\ell > 0$. As we see here, the long range tail of the potential leads to the exponential growth of $\varphi$ at $r\to \infty$, $\varphi \sim r^{1/4} \exp (2 \sqrt{\alpha r})$, and the exponential decrease of $\chi (r) \sim r^{1/4} \exp (- 2 \sqrt{\alpha r})$, to zero. This does not affect the validity of the change of variable $r \to x = \varphi/\chi$, etc, of section II, except that now $x$ grows exponentially as $r \to \infty$. Note that $I_1$ and $K_1$ do not vanish, $I_1$ for $r > 0$, and $K_1$ for $r < \infty$ \cite{8r}. $\varphi$ is again an increasing convex function, and $\chi$ a decreasing convex function. The only difference with the short-range potentials is that, now, $\chi^{-4}(r)$ grows exponentially, so that, in (\ref{49e}) and (\ref{64e}), if $V_0$ is chosen to be the Coulomb potential, the second potential may seem to become infinite as $r\to \infty$ ($x \to \infty$). However, we have always assumed $V_1(r)$ to be short range, i.e. decreasing fast enough at infinity. It follows that $\chi_0^{-4}(r) V_1(x)$ is again short range. For instance, if $V_1 (r) \sim r^{-4}$, then $\chi_0^{-4} V_1(r) \sim 1/(x^{2 } {\rm Log}^2 x)$, etc, i.e. $x\chi_0^{-4} V_1$ is $L^1$ in $x$ at $x= \infty$. \par

Other long range potentials of the form $g/r^n$, $n < 2$, can be dealt with in the same way by adapting (72) to $n < 2$, and one reaches similar conclusions as for the Coulomb potential. As for confining potentials like the harmonic oscillator, etc, we shall consider them in a separate paper.\\

\noi {\bf Concluding remarks.}\\

So far, all the potentials for which the radial Schr\"odinger equation
has been shown to be soluble analytically in closed form have their
solutions given by various hypergeometric functions in appropriate
variables \cite{1r,3r,4r,6r}. In fact, in many instances, as we have
seen on examples, the hypergeometric functions simplify to Bessel and
Hankel functions of real or imaginary arguments. The only exceptions
are Bargmann potentials \cite{1r,3r}, for which the solutions are given
in terms of rational functions of $\sinh \alpha_j r$, $\cosh \alpha_j
r$, $\sin \beta_j r$, and $\cos \beta_j r$, $j = 1, \cdots n$, where
$\alpha_j$ and $\beta_j$ are given by the positions of the poles and
zeros of the $S$-matrix, and the $r^{-4}$ potential \cite{2r}, for
which the solution is given in terms of Mathieu functions. \par

In the present paper, as seen on (\ref{49e}) and (\ref{64e}), the
potentials themselves have their arguments given by ratios of
hypergeometric functions, and the solutions are then hypergeometric
functions of ratios of hypergeometric functions, as seen on
(\ref{51e}). And this process can be repeated indefinitely, as we saw
before. One may then ask what the potentials and their wave functions
become in the limit. Also, we saw that, for both $V_0$ and $V_1$, we
can take potentials which are very singular but repulsive at the
origin, like $g r^{-n}$ and $\lambda r^{-m}$, $g > 0$, $\lambda > 0$,
$m, n > 2$. All kinds of combinations are therefore possible for $V_0$
and $V_1$.\par

It is trivial to construct infinitely many potentials for which the Schr\"odinger equation could be solved at zero energy. One can choose any positive, convex, and twice continously differentiable function, which is decreasing, and such that

$$\chi_0 (0) = 1 \ , \quad \chi_0 (\infty ) = {1 \over A}\ , \quad 1 < A < \infty \ , \eqno(77)$$

\noi and write

$$V_0 (r) = {\chi ''_0(r) \over \chi_0 (r)}\ . \eqno(78)$$

\noi If $\chi_0(r)$ is decreasing fast enough to $A^{-1}$ at infinity, then $V_0(r)$ is short-range. $\varphi_0(r)$ is defined here by $(13')$. Example :

$$\left \{ \begin{array}{l} \chi_0 (r) = \displaystyle{{1 \over (1 + \alpha )}} \left [ 1 + \displaystyle{{\alpha \over (1 + \beta r)^n}}\right ] \ , \\ \\ \alpha > 0\ , \ \beta > 0\ , \ n > 1 \ .\end{array} \right . \eqno(79)$$

\vskip 5 truemm
\noi According to (78), we have 

$$V_0(r) = { \alpha n (n+1) \over (1 + \alpha ) (1 + \beta r)^{n+2} \chi_0 (r)}\ \mathrel{\mathop \sim_{r \to \infty }} \ r^{-n-2}  \ . \eqno(80)$$

\vskip 5 truemm
\noi However, all this is valid for this particular $V_0(r)$. If we try to introduce a coupling constant $\lambda$ in front of $V_0(r)$, i.e. try to solve

$$\chi ''(r) = \lambda V_0(r) \ \chi (r)\ , \eqno(81)$$

\noi we usually do not find explicit solutions. This is indeed the case here. \par

Next, consider

$$\chi_0 (r) = {1 + e^{-\mu r} \over 2}\ , \ \mu > 0\ , \eqno(82)$$

\noi where $A=2$. This leads to

$$V_0(r) = {\chi '' _0 (r) \over \chi_0 (r)} = {\mu^2 e^{-\mu r} \over 1 + e^{-\mu r}}\ . \eqno(83)$$

\noi $\varphi_0 (r)$ is then easily calculated from ($13'$). In this case, one can solve (81) for any $\lambda$ since (83) is a particular case of (\ref{4e}), and the solutions are given, in general, in terms of hypergeometric functions \cite{4r}. Only for $\lambda = 1$, they reduce to the simple form (82) for $\chi_0$, and the corresponding $\varphi_0 (r)$.\\

\noi {\bf Acknowledgments.} One of the authors (KC) would like to thank the Department of Mathematics of the Science University of Tokyo, and Professors Kenro Furutani, Reido Kobayashi, and Takao Kobayashi, for warm hospitality and generous financial support when this work began.

\newpage
\begin{figure}
\centering
\includegraphics[width=8cm]{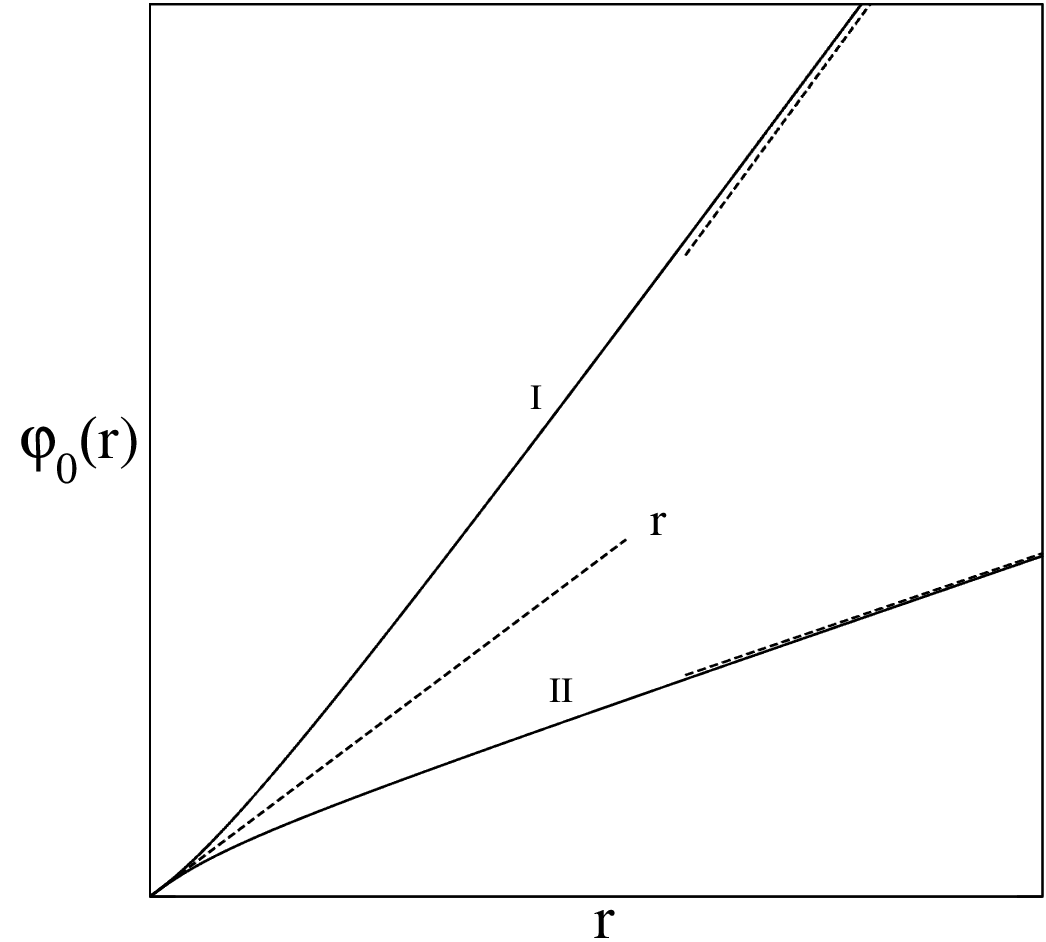}
\caption{$I, V > 0$ ; $II, V_0 < 0$, no bound states.}
%\end{figure}

%\begin{figure}
\centering
\includegraphics[width=8cm]{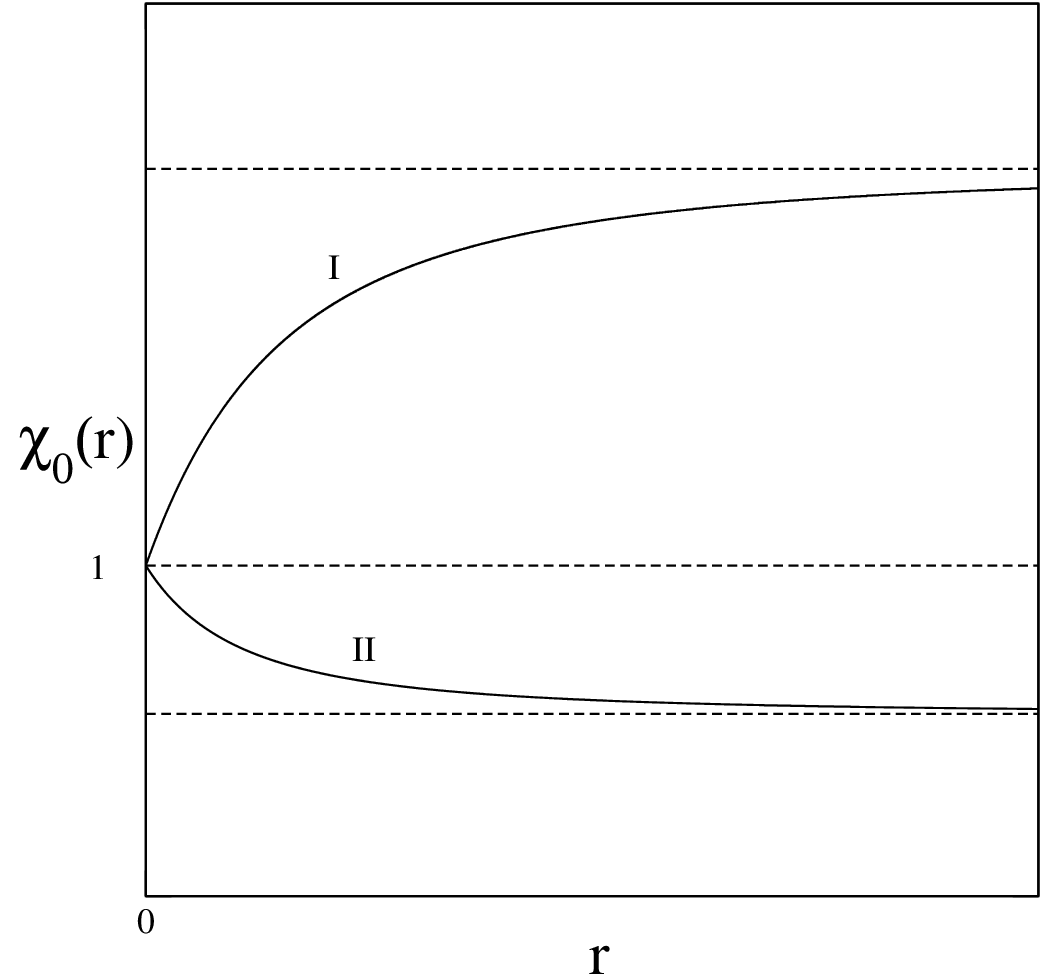}
\caption{$I,V_0 > 0$ ; $II, V_0 < 0$, no bound states.}
\end{figure}
\newpage
\begin{figure}
\centering
\includegraphics[width=8cm]{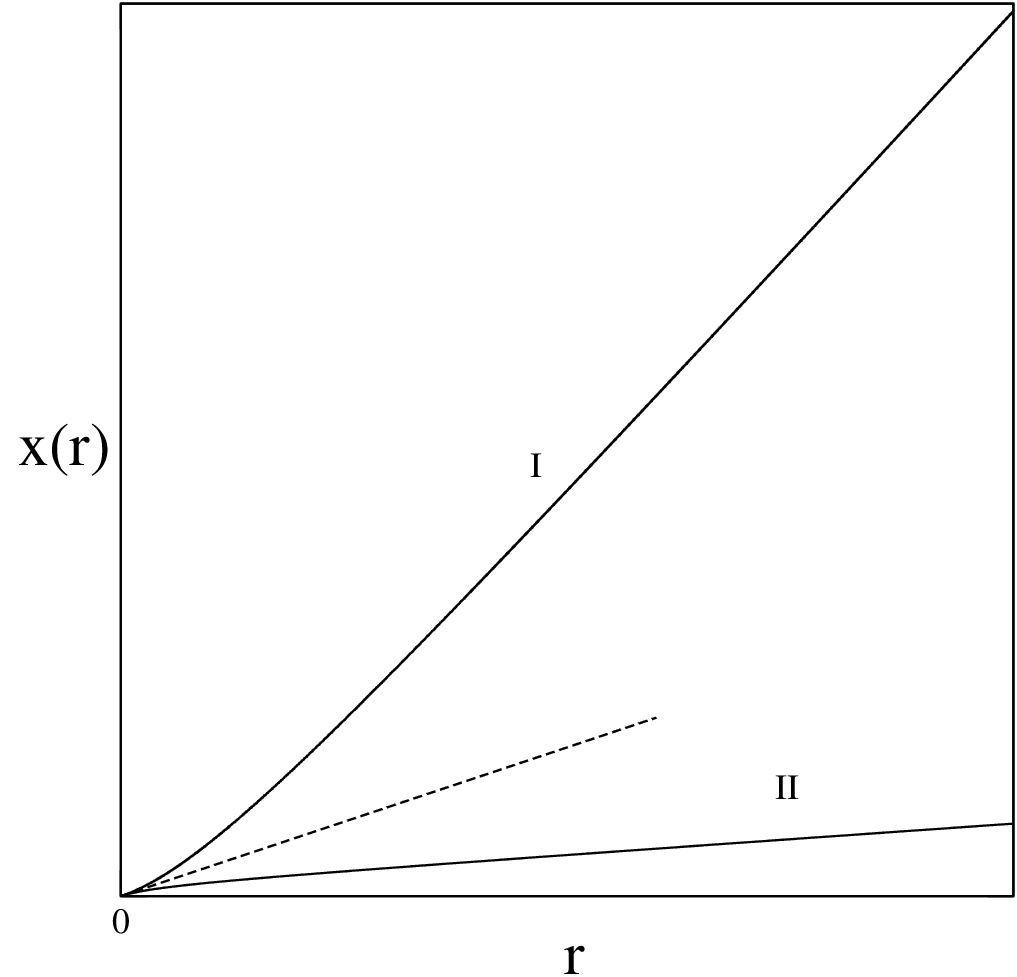}
\caption{$I, V_0 \geq  0$ ; $II, V_0 < 0$, no bound states.}
\end{figure}

\end{document}